\documentclass[a4paper,12pt]{article}
\pdfoutput=1
\usepackage{cite}
\usepackage{amsmath}
\usepackage{amsfonts}
\usepackage{amssymb}
\usepackage{graphicx, rotating}
\usepackage{epsfig}
\usepackage{latexsym}
\usepackage{graphicx}
\usepackage{color}
\usepackage{amsmath,bm,amssymb}
\usepackage{color}
\usepackage[english]{babel}
\usepackage{hyperref}

\newcommand{\Slash}[1]{{\ooalign{\hfil#1\hfil\crcr\raise.167ex\hbox{/}}}}

\newcommand{\beq}{\begin{equation}} \newcommand{\eeq}{\end{equation}}
\newcommand{\bef}{\begin{figure}} \newcommand{\eef}{\end{figure}}
\newcommand{\bec}{\begin{center}} \newcommand{\eec}{\end{center}}
 
\newcommand{\laq}[1]{\label{eq:#1}}  

\newcommand{\Eq}[1]{Eq.~(\ref{eq:#1})}

\def\({\left(}
\def\){\right)}

\def\O{\mathcal{O}}

\newcommand{\EV}{ {\rm \, eV} }

\newcommand{\MEV}{ {\rm \, MeV} }
\newcommand{\GEV}{ {\rm \, GeV} }
\newcommand{\TEV}{ {\rm \, TeV} }

\def\a{\alpha}

\def\d{\delta}

\def\f{\phi}
\def\g{\gamma}

\def\m{\mu}
\def\n{\nu}

\def\s{\sigma}

\def\G{\Gamma}

\def\L{\Lambda}

\def\tl{\tilde}
\def\*{\dagger}

\begin{document}

\begin{center}

\hfill  TU-1141\\

\vspace{1.5cm}
{\Large\bf Cosmological implications of $n_s\approx 1$ in light of the Hubble tension }
\vspace{1.5cm}

{\bf Fuminobu Takahashi$^{1,2}$} and {\bf Wen Yin$^{1}$}

\vspace{12pt}
\vspace{1.5cm}
{\em 

$^{1}$Department of Physics, Tohoku University,  
Sendai, Miyagi 980-8578, Japan \\
$^{2}$Kavli Institute for the Physics and Mathematics of the Universe (WPI),
University of Tokyo, Kashiwa 277--8583, Japan \vspace{5pt}}

\vspace{1.5cm}
\abstract{
Recently, a low-$z$ measurement of the Hubble constant, $H_0 = 73.04 \pm 1.04 {\rm ~km/s/Mpc}$, was reported by the SH0ES Team. The long-standing Hubble tension, i.e. the difference between the Hubble constant from the local measurements and that inferred from the
cosmic microwave background data
based on the $\L$CDM model, was further strengthened.
 There are many cosmological models modifying the 
 cosmology after and around the recombination era to alleviate this tension. 
 In fact, some of the models alter the small-scale fluctuation amplitude relative to larger scales, and thus require a significant modification of the primordial density perturbation, especially the scalar spectral index, $n_s$. 
 In certain promising models, $n_s$ is favored to be larger than the $\L$CDM prediction, and even the scale-invariant one, $n_s=1$, is allowed. In this Letter, 
 we focus on the very early Universe models to study the implication of such unusual $n_s$. 
 In particular, we find that an axiverse with an axion in the equilibrium distribution during inflation can be easily consistent with $n_s = 1$. This is because the axion behaves as a curvaton with mass much smaller than the inflationary Hubble parameter. 
 We also discuss other explanations of $n_s$ different from that obtained based on the $\Lambda$CDM.
}
\end{center}
\clearpage

\setcounter{page}{1}
\setcounter{footnote}{0}

\section*{}
{\bf Hubble tension and $n_s$.--}
The Hubble constant ($H_0$) inferred from the $\L$CDM model with the input of the temperature/polarization anisotropies of the cosmic microwave background (CMB) is \cite{Planck:2018vyg}
\beq 
H_0^{\rm CMB}= 67.4 \pm 0.5 {\rm ~km ~s^{-1}~Mpc^{-1}}.
\eeq 
This is known to differ from the locally 
measured value at low red shifts
\cite{Riess:2011yx, Bonvin:2016crt,Riess:2016jrr, Riess:2018byc,Birrer:2018vtm} by 
at least $4\s$ level (see reviews ~\cite{Freedman:2017yms, Verde:2019ivm,DiValentino:2020zio,Schoneberg:2021qvd}.)
Recently, the SH0ES Team has reported a new result from the measurement of 42 SNe la at $z<0.01$, in which  
\cite{Riess:2021jrx}
\beq 
H_0 = 73.04 \pm 1.04 {\rm ~km ~s^{-1}~Mpc^{-1}}.
\eeq 
Thus the long-standing discrepancy is further strengthened and the new result shows an around $5\s$ difference from the $\L$CDM prediction. 

There have been proposed many models altering the cosmology around or after the recombination era
to relax the Hubble tension (see e.g. \cite{Schoneberg:2021qvd} and the references therein). 
In particular, some of them require different initial conditions for the very early Universe from the $\L$CDM model. 
For instance, the primordial curvature perturbation is given in the form of
\beq
{\cal P}_{\zeta}[k]={\cal P}_{\zeta}[k_*] (k/k_*)^{n_s-1}. 
\eeq
where the index $*$ denotes the variable evaluated at the horizon exit of the CMB scales here and hereafter, and $n_s$ is the scalar spectral index.
 In various solutions to the Hubble tension, $n_s$ is often diffenrent from the $\Lambda$CDM value, \beq n_s \neq n_s^{\rm \L CDM} 
 \eeq
with
$ 
n^{\L\rm CDM}_s= 0.965\pm 0.004
$~\cite{Planck:2018vyg}. Intriguingly,
in the light of the Hubble tension $n_s=1$, is not ruled out~\cite{DiValentino:2018zjj,Ye:2021nej}. 
 
Moreover, certain promising scenarios predict a relative suppression of the small scale fluctuations. This can be compensated by increasing $n_s$. Then, $n_s> n_s^{\L\rm CDM}$ is even favored. For instance, in the early dark energy scenario and its modifications~\cite{Karwal:2016vyq, Poulin:2018cxd, Poulin:2018dzj,Lin:2019qug,Niedermann:2019olb,Niedermann:2020dwg, Murgia:2020ryi, Smith:2020rxx},
an enhanced early integrated Sachs-Wolfe effect suppresses the growth of perturbations which leads to a larger $n_s$ and a higher density of cold dark matter
\cite{Ye:2021nej,Vagnozzi:2021gjh}.\footnote{A scalar field that contributes to the early dark energy may also be indicated from the recent measurement of the isotropic cosmic birefringence~\cite{Minami:2020odp}. See also the explanations by slow-rolling axion-like particle~\cite{Minami:2020odp,Fujita:2020aqt} or domain wall without a string ~\cite{Takahashi:2020tqv} and its possible relation to the Hubble tension~\cite{Capparelli:2019rtn, Fujita:2020aqt,Reig:2021ipa}.}. Another example is the $\EV $ scale majoron which was proposed to explain the Hubble tension~\cite{Escudero:2019gvw}. 
The hot majoron around the 
recombination era seems to suppress 
the small scale structure which gives a preference to $n_s>n_s^{\rm \L CDM}$ from the 
$\L$CDM value.\footnote{On the contrary if the majoron is parametrically heavy, we obtain self-interacting neutrinos in the effective theory. This predicts a slightly lower $n_s$ than the $\L$CDM one because of the presence of the non-free-streaming neutrino component, and the small-scale fluctuation is less suppressed than the $\L$CDM case~\cite{Cyr-Racine:2013jua, Lancaster:2017ksf, Kreisch:2019yzn}. Such scenarios can be consistent with certain inflation models. See the last part of the Letter.}
There are some other solutions to the Hubble tension which also 
have the best fit of $n_s\approx 1$ e.g.~\cite{Lin:2019qug,Niedermann:2020dwg, Murgia:2020ryi,Escudero:2019gvw}. 

Given those observations and the strengthened significance of the Hubble tension, it may be a good point to discuss their implications for the very early Universe. In this Letter, we discuss some simple cosmological models 
to increase or decrease $n_s$ than the $\L$CDM value. In particular, we will mainly focus on the $n_s\approx 1$ cosmology which may be preferred from various scenarios for explaining the Hubble tension.  

{\bf Inflation with $n_s \approx 1$.--}
The slow-roll inflation solves various fine-tuning problems associated with the initial condition of the big-bang cosmology. More importantly, the quantum fluctuations of the inflaton are known to generate the density perturbations. The predicted scalar spectral index in the simple slow-roll inflation models is
\begin{equation}
  n_s-1 \simeq  - 6 \varepsilon + 2 \eta,
\end{equation}
where $\varepsilon$ and $\eta$ are the slow-roll parameters of the inflaton. Since $n_s$ depends on the functional form of the inflaton potential as well as thermal history after inflation, it is often used to constrain inflation models. On the other hand, $n_s$ is known to be rather sensitive to tiny corrections to the inflaton potential, especially in the so-called small-field inflation. For instance, the predicted $n_s$ is smaller than $n^{\L\rm CDM}_s$ in the quartic hilltop inflation, but it can be easily enhanced close to or even larger than $n^{\L\rm CDM}_s$ by introducing a tiny linear term~\cite{Takahashi:2013cxa}. In many inflation models which predict $n_s < 1$ in their original set-up, it is similarly possible to add some modifications to the inflaton potential to realize $n_s \approx 1$. This is an interesting direction to study. However, here we instead pursue a scenario that generically predicts $n_s \approx 1$ over wider parameter space.

{\bf Review on curvaton.--}
Let us briefly review the curvaton scenario~\cite{Linde:1996gt, Enqvist:2001zp,Lyth:2001nq, Moroi:2001ct} in which the curvature perturbation is generated by the fluctuation of a curvaton field $\s$ during inflation rather than the inflaton fluctuation.
We assume that $\s$ does not have a significant interaction to the inflaton, nor a non-minimal coupling to gravity. Let us decompose the curvaton field into
the spatially homogeneous part and a fluctuation about it, $\s[\vec{x}] =\s_0 +\d \s[\vec{x}]$,
and denote 
the Fourier component of the perturbation as $\d \s_{\vec{k}}$, where the time dependence 
is not shown explicitly. 
The fluctuation evolves following
\beq
\d \ddot\s_{\vec{k}}+3 H \d \dot \s_{\vec{k}} +((k/a)^2+V'') \d \s_{\vec{k}}=0,
\eeq
where $a$ is the scale factor, $H$ is the Hubble parameter,
$V=V[\s]$ is the curvaton potential, and the prime denotes the derivative with respect to the variable of the function.
When $|V''| \lesssim H^2_*$, the fluctuation of the curvaton generated during inflation is given by
\beq
(\d \s_{\vec{k}})^2\simeq \(\frac{ H_{*}}{2\pi}\)^2(k/k_*)^{(n_\s-1)}
\eeq
with 
\beq
n_\s-1 = \(2 \frac{\dot{H}}{H^2} +\frac{2}{3} \frac{V''}{H^2}\)_*.
\eeq
In the slow-roll inflation,
$\dot{H}/H^2$ can be sizable only for inflation models with super-Planckian field excursion, and such inflation models are often assumed to realize $n_s < 1$ in the context of curvaton scenario. However, a super-Planckian field excursion is known to be in tension with the quantum gravity. Thus, in the following we neglect the first term, and focus on the second term in the parenthesis.

The curvaton $\s$ starts to oscillate when the mass becomes comparable to the Hubble parameter after inflation.
The perturbation in the energy density is obtained as
\beq
\frac{\d \rho_{\s}}{\rho_{\s}}\approx \frac{\sqrt{(\d \sigma_{\vec{k}})^2}}{\sigma_0 }\simeq \frac{2 H_*}{2\pi \sigma_0}
\eeq
by assuming $H_* \ll \s_0$, and $V \propto \sigma^2 $ at the onset of the oscillation.\footnote{If $V\propto \sigma^n$ with $n>2$ the curvaton has a non-uniform onset of its oscillation. There will be additional contributions to the final density perturbations~\cite{Kawasaki:2011pd}, which we do not consider here.} 
We assume that the curvaton subsequently dominates over the Universe and it later decays to the standard model (SM) particles it couples to. The curvaton decay reheats the Universe and the SM particles make up the thermal bath of the big-bang cosmology. 
Then, the density perturbation of the plasma follows that of the curvaton, 
\beq
{\cal P}_{\zeta}\approx \(\frac{ H_*}{\pi \sigma_0}\)^2, ~~n_s\approx \left. 1+\frac{2}{3}\frac{V''}{H^2}\right|_*.
\eeq
Since the CMB normalization gives~\cite{Aghanim:2018eyx}
\beq\laq{CMBnorm}
{\cal P}_{\zeta}\approx 2\times 10^{-9},
\eeq
we obtain
\beq
\laq{Hubbleamp}
H_* \approx 1.4\times 10^{12}\GEV \left(\frac{\sigma_0}{10^{16}\GEV } \right).
\eeq
A bound on $H_*$ is derived from the tensor-to-scalar ratio~\cite{BICEP:2021xfz}
\beq
\laq{ttos}
r\equiv {\cal P}_h/{\cal P}_\zeta <0.036
\eeq
which together with \Eq{CMBnorm} gives an upper limit on the tensor mode. This bound leads to
\beq
H_{*}\lesssim 4.7\times 10^{13}\,\GEV.
\eeq 
This upper bound on the inflation scale with the CMB normalization implies that the homogeneous part of the curvaton field, $\sigma_0$, is sub-Planckian. 
We emphasize that in our scenario the inflation is not required to explain the density perturbation and it is quite easy to build such models. Thus in the following, we do not specify an inflation model, and only require that the inflaton field excursion is sub-Planckian.

In the context of curvaton, one often considers a sizable tachyonic mass $\sqrt{|V''|} \sim 0.1 H_*$ to explain $n_s=n_s^{\rm \L CDM} < 1$. 
However, in this case, the curvaton starts to oscillate soon after inflation, and it is difficult to dominate the Universe unless the initial position is close to the top of the potential~\cite{Kawasaki:2011pd,Kawasaki:2012gg}. Also, in this case, the non-Gaussianity is known to be logarithmically enhanced. 
If we allow the curvaton potential to be time-dependent, 
 a natural way may be to introduce a non-minimal coupling of ${\cal L}\supset \xi \sigma^2 R $ with $|\xi| = {\cal O}(0.01)$ and then $V''$ in the Einstein frame gets a contribution of $\O(\xi H_*^2)$.

{\bf Curvaton scenarios in light of Hubble tension.--}
As we mentioned, in this Letter we focus on $n_s \approx 1$. This implies that, in the context of a curvaton scenario, 
\beq\laq{ns0}
|(V'')_*| \ll H^2_*
\eeq
for which $n_s\approx 1$ is naturally realized. 
The small curvature of the potential also means the suppression in the non-minimal coupling. 
From the viewpoint of naturalness, the suppression may imply symmetry, e.g. supersymmetry or shift symmetry. In the following, we consider the latter to suppress the curvaton mass.

{\bf ALP as curvaton.--}
An interesting possibility is that the curvaton is an axion~\cite{Enqvist:2001zp, Dimopoulos:2003az, Kawasaki:2012gg}. 
In this case the potential naturally has the form of
\beq
V=\L^4 (1-\cos[\s /f_a])
\eeq
 where $\L$ is a dynamical scale that breaks the ``Peccei-Quinn" symmetry, and $f_a$ is the decay constant. We assume $\L$ does not depend on the cosmic time. 
 The mass, $m_a$, of the axion is obtained as
 \beq
 m_a= \frac{\L^2}{f_a}. 
 \eeq

For successful reheating, we need to couple the axion to the SM particles.  
In particular, an axion coupled to photons is called an axion-like-particle (ALP).
The ALP coupling to a pair of photons is given by
\begin{align}
\laq{int}
{\cal L} = c_\g \frac{\a}{4 \pi} \frac{\s}{f_a} F_{\mu \nu} \tilde F^{\mu \nu} 
\end{align} 
where $c_\g$ is the anomaly coefficient, $\a$ the fine-structure constant, and $F_{\m\n}$ ($\tl F_{\m\n}$) the photon field strength (its dual), 
 Then the axion or ALP 
 can decay into a photon pair with a decay rate of
 \footnote{More precisely, when the axion is heavier than the weak scale, we have to include the decay to the weak gauge bosons or some other particles, which may induce the photon coupling. Those effects generically raise $T_R$ but we do not consider those model-dependent effects.}
\beq
\Gamma_{\rm dec,\g} = c_\g ^2\frac{\alpha^2 }{64 \pi^3} \frac{m_{a}^3}{f_a^2}.
\eeq

During inflation, the light axion of the mass satisfying $m_a\ll H$ classically rolls towards the potential minimum. 
Let us assume that the inflation scale $H$ does not change much during inflation, $H\sim H_*$, here and hereafter. This is satisfied in various inflation models, especially, with a sub-Planckian field excursion. 
In addition to the classical motion, there is also quantum diffusion during inflation. 
The coarse-grained field ($(\d \s[\vec{k}])^2 $ averaged over the super-horizon modes) evolves following the so-called Fokker-Planck equation~\cite{Starobinsky:1986fx,Starobinsky:1994bd}
\begin{eqnarray}
\frac{\partial{P}(\sigma, t)}{\partial t}=\frac{1}{3H}\frac{\partial}{\partial \sigma}(V'{P}(\sigma, t))+\frac{H^3}{8\pi^2}\frac{\partial^2{P}(\sigma, t)}{\partial \sigma^2},
\end{eqnarray}
 For sufficiently long period of inflation, $t\to \infty,$
the axion naturally follows the equilibrium distribution, when the classical motion balances the quantum diffusion~\cite{Starobinsky:1986fx,Starobinsky:1994bd,Hardwick:2017fjo, Graham:2018jyp,Guth:2018hsa,Nakagawa:2020eeg}.  
When the axion follows the equilibrium distribution during inflation, we get the probability distribution of 
\beq
\laq{dist}
P[\sigma_0]\propto \exp{\(- \frac{8\pi^2}{3 H^4}{V[\sigma_0]}\)}.
\eeq
Such a stochastic axion curvaton is an interesting possibility to explain $n_s \approx 1$, as we shall see shortly.

The curvaton in a stochastic set-up was discussed in Ref.\,
\cite{Hardwick:2017fjo}, in which they studied the equilibrium distribution of a spectator field focusing on whether the curvaton distribution follows the equilibrium one toward the end of inflation. They found that the equilibrium distribution is not kept in the chaotic type inflation. 
In the case of the axionic spectator at the end of inflation, the authors set the spectator field mass to be around $10\%$ of the Hubble parameter. 
In contrast, our focus is $n_s\approx 1$, and $m_a \ll H_*$. In this case, the $\s$ distribution reaches equilibrium during the very long inflation with the Hubble rate of $H\sim H_*$,
much before the horizon exit of the CMB scales.
The early equilibrium distribution is kept intact until the end of the inflation. 

There are two limits: $ H_*\ll \L$ and $H_*\gg \L$. 
In the former case 
we can expand $V$ around the minimum, and adopt a quadratic approximation. 
Then, the typical field value of the axion during inflation is obtained from \Eq{dist}
\beq
\s_0 \sim \s_{\rm stochastic}\equiv \sqrt\frac{3 H_*^4}{8\pi^2 m_a^2}
\eeq 
From \Eq{Hubbleamp} we get
\beq
H_* \sim 4\times 10^4\, m_a ~~~~~(H_*\ll \L).
\eeq
This condition satisfies \Eq{ns0}.
More precisely, the contribution to the spectral index is \beq
\delta{n_s}\sim 10^{-7}.
\eeq

The conditions for the axion to be the curvaton are as follows. 
The axion reheats the Universe at the temperature estimated from $T_R\equiv \({\frac{g_{\star ,\rm dec}\pi^2}{90}}\)^{-1/4}\sqrt{M_{\rm pl} \G_{\rm dec,\g}}$ 
where $g_{\star ,\rm dec}$ denotes the relativistic degrees of freedoms at the decay.\footnote{There is also the dissipation effect~\cite{ Moroi:2014mqa} $\G \propto m_a^2 \frac{T}{f_a^2}$ if the temperature of the Universe at the onset of the oscillation of the ALP is high.
As one can see, this becomes important at a lower temperature in either matter or radiation dominated Universe. Therefore we can compare it with the Hubble rate when $T\sim m_a$. At this moment, the dissipation is subdominant compared with the decay. 
} 
The reheating temperature should be higher than MeV so that the big-bang nucleosynthesis (BBN) is not spoiled~\cite{Kawasaki:1999na,Kawasaki:2000en,Ichikawa:2005vw}. This reads
\beq
T_R\approx 0.02\GEV c_\g \(\frac{g_{\star,\rm dec}}{10}\)^{-1/4} \(\frac{m_a}{10^6\GEV}\)^{3/2}\(\frac{10^{16}\GEV }{f_a}\) > \MEV.
\eeq

\begin{figure}[!t]
\begin{center}  
\includegraphics[width=110mm]{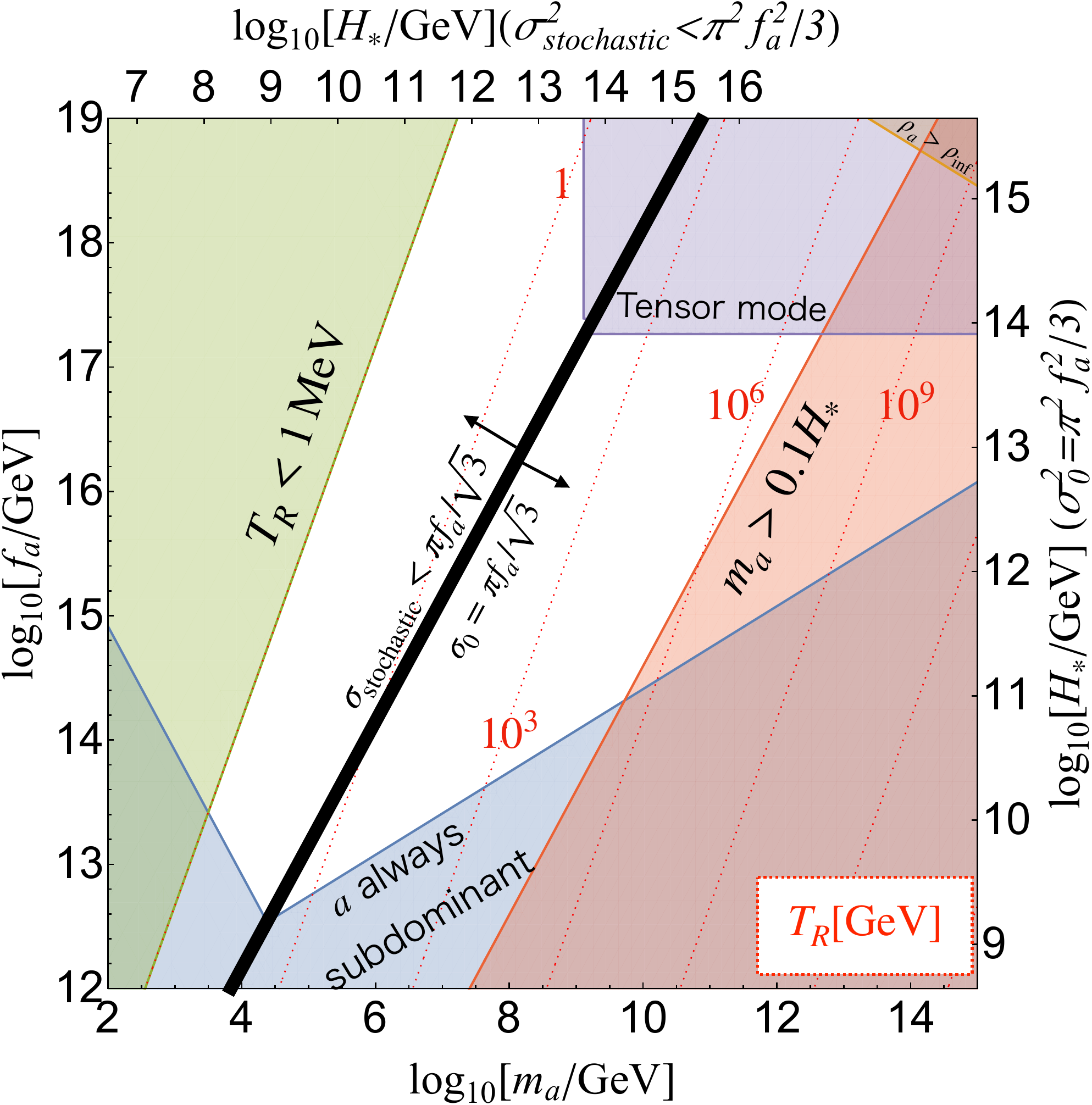}   
  \end{center}
\caption{ The parameter region in $m_a-f_a$ plane with the contours of the reheating temperature, $T_R$, caused by the ALP decay. On the upper (right) axis, we show $H_*$ corresponding to the region left (right) to the black solid line. The left green region is excluded due to the too low reheating temperature which is incompatible with the BBN. The blue region in the bottom does not have a period of the curvaton dominated Universe. In the upper-right purple (orange) region, the tensor mode is too large (the axion-energy dominates over the inflation energy density). 
In the right red region, $m_a>H_*$ and the axion-curvaton oscillation starts during inflation. In this figure $c_\g=1$ is taken. 
}\label{fig:1} 
\end{figure}

Another important condition is that the axion must once dominate the Universe.\footnote{Strictly speaking, even if the axion curvaton is subdominant, one can increase $(H_*/\sigma_0)^2$ to enhance the perturbation but the non-Gaussianity tends to be too large~\cite{Planck:2019kim}.}
Let us assume for simplicity that the onset of oscillation occurs in the radiation dominated epoch.\footnote{The reheating temperature of the inflaton decay should not be too high, otherwise the ALP is also thermally produced in some parameter region. Since such thermally produced ALPs carry the fluctuation from the inflaton field, the energy contribution should be sub-dominant. This constraint further narrows the parameter region to $f_a\gtrsim 10^{14}\GEV,m_a\gtrsim 10^6\GEV $ in Fig.\ref{fig:1}. That said, when the reheating by the inflaton completes just before the onset of oscillation, this constraint does not apply.} From the standard misalignment mechanism, 
we obtain the would-be abundance by neglecting the axion decay~(the numerical fit of the coefficient is taken from Ref.\,\cite{Ho:2019ayl})
$
\Omega^{ a}_{\phi} h^2 
\,\sim\,
 10^{15} 
\bigg(\frac{g_{\star,\text{osc}}}{106.75}\bigg)^{-1/4}
\bigg(\frac{m_a}{10^6\,\text{GeV}}\bigg)^{1/2} 
\bigg(\frac{\s_\text{0}}{10^{16}\,\text{GeV}}\bigg)^2, 
$
where $h$ is the reduced Hubble constant, and $g_{\star,\text{osc}}$ is the degrees of freedom at the onset of oscillation. 
We can estimate the temperature, $T_{\rm dom},$ that the axion dominates the Universe, from 
$
\Omega^{\rm stable}_{a} \frac{\rho_c}{ s_0 } \times \frac{3}{4T_{\rm dom}}=1
$
with $s_0, \rho_c$ being the present entropy density and the critical density. 
This gives 
\beq
T_{\rm dom}\approx 2\TEV \(\frac{m_{a}}{10^6\GEV}\)^{5/2}.
\eeq
One arrives at the condition that the axion domination happens:
\beq 
T_{\rm dom}>T_R
\eeq

When $H_* \gg \L$, the exponent of \Eq{dist} is always much smaller than unity $1$. We have an almost flat distribution of the axion. In this case we take $\sigma_0 = \pi f_a/\sqrt{3}$. Then we find 
\beq
H_* \sim 3\times 10^{12}\GEV \left(\frac{f_a}{10^{16}\GEV}\right)~~~~~(H_*\gg \L).
\eeq
The conditions to be satisfied are quite similar to the previous case, and we do not repeat them here.

The viable parameter region in $m_a-f_a$ plane is shown in Fig.\ref{fig:1}, with the contour of the reheating temperature. From the right top colored regions are excluded due to the non-observation of the tensor mode, $|n_s-1| \gtrsim 0.01$ due to $m_a>0.1H_*$, subdominant axion until its decay, and the BBN bound in the clockwise direction. In the small orange region in the top-right corner, the ALP energy density during inflation $\rho_a \equiv \L^4$, is larger than the inflationary energy density $ \rho_{\rm inf}\equiv 3H_*^2 M_{\rm pl}^2$.
Much left (right) to the black solid line has $\L \gg H_*$ ($\L\ll H_*$) where the corresponding $H_*$ is shown in the upper (right) axis. 
The viable parameter region is in the range 
\beq
100\GEV<m_a< 10^{12}\GEV, 10^{12}\GEV<f_a\lesssim M_{\rm pl}.
\eeq
The corresponding inflationary Hubble parameter is 
\beq
10^{8}\GEV<H_*< 4.7 \times 10^{13}\GEV.
\eeq
The reheating temperature due to the curvaton decay is in the range
\beq
1\MEV \lesssim T_R\lesssim 10^7\GEV.
\eeq
Therefore various baryogenesis scenarios and dark matter productions are comparable.\footnote{Around the low reheating regime, $m_a$ is heavy, and also the baryogenesis due to the axion decay is possible given a proper coupling of the axion together with certain BSM particles. }
Much before the decay of the axion, the inflaton or the plasma produced from the inflaton may dominate the Universe. In particular, around the blue bottom region, such component is non-negligible even at the reheating by the axion decay. Nevertheless, the baryogenesis or dark matter production should happen relevant to the curvaton $\s$ but not the inflaton, because otherwise the baryon or dark matter has ``isocurvature".

In most of the parameter range, we have $n_s\approx 1$, which is quite a natural prediction in this scenario.  
On the other hand, around the boundary of the right red region denoting $m_a>0.1H_*$,  
 $|n_s-1|\sim 0.01$ can be realized. In this case, whether $n_s$ is greater or smaller than $1$ is determined by the sign of the 
 curvature $V''$ of the potential, which is approximately a 50/50 chance.

{\bf Cosmology in axiverse with $n_s\approx 1$.--} 
So far, we have shown that the ALP-curvaton scenario is consistent with those solutions to the Hubble tension that predict $n_s\approx 1$.
The resulting decay constant is consistent with the one for the conventional string axion~$f_a\sim 10^{15-17}\,$GeV. 

This may imply that the Universe is in the axiverse~\cite{Witten:1984dg, Svrcek:2006yi,Conlon:2006tq,Arvanitaki:2009fg,Acharya:2010zx, Higaki:2011me, Cicoli:2012sz,Demirtas:2018akl,Mehta:2020kwu,Mehta:2021pwf}, 
in which there are many axions with masses spread over a wide mass range.
In particular, an ALP coupled to the SM gauge bosons is predicted in an M-theory compactification~\cite{Acharya:2010zx}. 
In this scenario, it is natural that a string axion happens to have the mass in the viable range of Fig.\ref{fig:1} and plays the role of the curvaton. In this case, the predicting Universe naturally has $n_s\approx 1$. 

We comment on a possible relevant cosmology in the axiverse with the stochastic axion curvaton. 
Lighter axions, if exist, also have the stochastic nature~\cite{Graham:2018jyp,Guth:2018hsa}. In the usual single cosine potential with a potential hight $\sim \L_{\rm DM}^4$ satisfying $\L_{\rm DM}\gg H_{*}$ a lighter axion tends to dominate the Universe. For the viable inflation scales in Fig.\ref{fig:1}, the natural mass scale of the axion to explain the dark matter abundance 
is too heavy to be a stable ALP (see Ref.~\cite{Ho:2019ayl}). In most of the sub-MeV mass range where the ALP is stable and can evade observational constraints, we need to tune the initial amplitude of the ALP in the equilibrium distribution. 
A special mass range without tuning is $\sim 10^{-18}\EV$ for a decay constant $\sim 10^{16}\GEV$. In this case, the equilibrium distribution is flat and the initial misalignment amplitude is around the decay constant.

With multiple cosine terms which make up a flat-bottomed potential, on the other hand, 
the axion dark matter can be even lighter without over-closing the Universe \cite{ Daido:2016tsj, Nakagawa:2020eeg}.
In this case, $H_{*}\lesssim 10^{13}\GEV$ is further needed to evade the isocurvature bound~\cite{Nakagawa:2020eeg}.

If a very light axion, whose mass is such light that its energy contribution to the Universe is subdominant, may start its oscillation around or after the recombination. 
If the light axion gets this kind of potential, the Hubble tension can be relaxed via the early dark energy which is transferred into the axion oscillation energy which scales as radiation~\cite{Poulin:2018cxd, Poulin:2018dzj}. The initial condition of the axion can be set by the stochastic behavior during inflation. In this case, there can also be a domain wall formation depending on whether the initially averaged axion is close to the potential top~\cite{Khlopov:2004sc, Takahashi:2020tqv}.  
The domain wall problem can be evaded since the axion is light.

In addition, the inflation itself may be caused by an axion in the axiverse~\cite{Freese:1990rb,Adams:1992bn, Czerny:2014wza, Czerny:2014xja,Czerny:2014qqa,Higaki:2014sja, Croon:2014dma,Higaki:2015kta, Higaki:2016ydn}.
In particular, with a flat axion potential top composed by several-cosine terms, 
 the inflation can happen with a sub-Planckian decay constant~\cite{Czerny:2014wza,Czerny:2014qqa,Croon:2014dma,Higaki:2015kta}.
The flat top, as well as the flat bottom, can be realized in extra-dimensional models~\cite{Croon:2014dma,Higaki:2015kta, Higaki:2016ydn}. It is intriguing that such inflaton potential resembles that used for the early dark energy, both of which may be
realized in the axiverse or the axion landscape.

{\bf QCD axion as curvaton.-- } 
A natural question is whether the QCD axion~\cite{Peccei:1977hh,Peccei:1977ur,Weinberg:1977ma,Wilczek:1977pj} can be the curvaton for our purpose. Naively the answer seems to be negative because of the smallness of the QCD scale. 
Here we discuss the possibility that the QCD axion acquires an aligned instanton contribution from UV dynamics to the IR QCD instanton. 
In this case, the QCD axion can be much heavier than the conventional one, while the strong CP problem is solved. To be specific, we consider a mirror SM sector related to the SM by $Z_2$ symmetry following \cite{Takahashi:2021tff} (see also the original proposal in Ref.\cite{Fukuda:2015ana}). We assume that the $Z_2$ symmetry is broken by a VEV of hidden Higgs, $v_H$. $v_H$ is much larger than the electroweak (EW) scale.

We consider the (string) axion couplings to gluons and hidden gluons (shown with primes) as follows,
\beq
-{\cal L} = \frac{\a_s}{8\pi}( \frac{\s }{f_a} -\theta_{CP}) G\tl G +\frac{\a'_s}{8\pi}( \frac{\s }{f_\a} -\theta_{CP}) G'\tl G'. 
\eeq 
at a renormalization scale, $\m_{\rm RG} (<v_H)$. At this scale, both sectors are still perturbative.
While $\a_s' \sim \a_s$ at $\mu_{\rm RG}\sim v_H$, we have $\a_s'> \a_s$ at $\mu_{\rm RG} \ll v_H$ due to the decoupling of the heavy colored particles in the hidden sector. The hierarchy between $\a_s'$ and $\a_s$ can be even larger with extra colored particles in both sectors. Those particles get masses around $v_H$ in the hidden sector while their $Z_2$ partners
in the SM sector remain relatively light but not too light to be consistent with the experimental bounds.

Then, non-perturbative effects of the hidden QCD provides a large contribution to the axion potential whose minimum is aligned to the vanishing strong CP phase, 
\beq
V\sim 
-\Lambda'^4
\cos {(\frac{\s}{f_a}-\theta_{CP})},
\eeq
Note that in the hidden sector the axion gets mass from a pure Yang-Mills instanton due to the decoupling of the hidden quarks. 
In this case, the axion may not have sizable couplings to the hidden photon unlike the standard QCD axion which mixes with mesons.  
The dominant decay of the axion is to the SM gluons and it reheats the Universe. This scenario gives a similar parameter region as in Fig.\ref{fig:1}.\footnote{If inflaton reheats the hidden gluon so that the axion potential vanishes due to the thermal effect, however, our discussion may change. Since the axion is easier to dominate the Universe, the lower bound of the decay constant may be smaller. One should also consider the evaporation of the axion condensate due to the dark QCD spharelon effect. }

{\bf Other possibilities of explaining $n_s$ different from $n_s^{\rm \L CDM}$ .--} When $n_s$ is close to, but not very close to 1, the
chaotic inflation with a monomial potential $V_{\rm inf}\propto \f^p $ with small $p$ is a simple way to realize such $n_s$ together with a suppressed tensor mode~\cite{BICEP:2021xfz} (see also \cite{Akrami:2018odb}). 
For a sufficiently small $p (>0)$, the predicted spectral index asymptotes to $n_s\simeq 0.98$ for the e-folding number $N = 50$.
However, we need a super-Planckian field excursion, which may be disfavored from the quantum gravity. 
The multi-natural inflation~\cite{Czerny:2014wza,Czerny:2014qqa,Croon:2014dma,Higaki:2015kta}, ALP inflation~\cite{Daido:2017wwb, Daido:2017tbr, Takahashi:2019qmh, Takahashi:2019pqf}, or the QCD axion inflation~\cite{Takahashi:2021tff} also have the parameter region of $n_s\sim 1$ (see also Ref.\cite{Takahashi:2013cxa} for the mechanism to modify $n_s$). 
The same models can explain $n_s<n_s^{\rm \L CDM}$.

\section*{Acknowledgments}
This work is supported by JSPS KAKENHI Grant Numbers 17H02878 (F.T.), 20H01894 (F.T.), 20H05851 (F.T. and W.Y.), and 21K20364 (W.Y.)

\bibliography{reference}
\end{document}